# Demonstrating a Pre-Exascale, Cost-Effective Multi-Cloud Environment for Scientific Computing

Producing a fp32 ExaFLOP hour worth of IceCube simulation data in a single workday

Igor Sfiligoi
University of California   San Diego
La Jolla CA USA
isfiligoi@sdsc.edu

David Schultz
University of Wisconsin - Madison
Madison WI USA
dschultz@icecube.wisc.edu

Benedikt Riedel
University of Wisconsin - Madison
Madison WI USA
briedel@icecube.wisc.edu

Frank Wuerthwein
University of California San Diego
La Jolla CA USA
fkw@ucsd.edu

Steve Barnet
University of Wisconsin - Madison
Madison WI USA
barnet@icecube.wisc.edu

Vladimir Brik
University of Wisconsin - Madison
Madison WI USA
vbrik@icecube.wisc.edu

## ABSTRACT

Scientific computing needs are growing dramatically with time and are expanding in science domains that were previously not compute intensive. When compute workflows spike well in excess of the capacity of their local compute resource, capacity should be temporarily provisioned from somewhere else to both meet deadlines and to increase scientific output. Public Clouds have become an attractive option due to their ability to be provisioned with minimal advance notice. The available capacity of cost-effective instances is not well understood. This paper presents expanding the IceCube's production HTCondor pool using cost-effective GPU instances in preemptible mode gathered from the three major Cloud providers, namely Amazon Web Services, Microsoft Azure and the Google Cloud Platform. Using this setup, we sustained for a whole workday about 15k GPUs, corresponding to around 170 PFLOP32s, integrating over one EFLOP32 hour worth of science output for a price tag of about $60k. In this paper, we provide the reasoning behind Cloud instance selection, a description of the setup and an analysis of the provisioned resources, as well as a short description of the actual science output of the exercise.

## CCS CONCEPTS

• Computer systems organization---Distributed architectures---Cloud computing;500 • Software organization and properties---Software system structures---Distributed systems organizing principles;300   • Applied computing---Physical sciences and engineering;100

## 1 Introduction

Scientific computing needs are growing dramatically with time and many communities have to occasionally deal that workloads that exceed the capacity of their local compute resource. At the same time, public Cloud computing has been gaining traction, including funding agencies starting to invest in this sector; examples being NSF's E-CAS and CloudBank awards, and the European Cloud Initiative. Cloud computing, with its promise of elasticity, is the ideal platform for accommodating occasional spikes in computing needs. In a past exercise [1] we demonstrated that it is possible to provision and effectively use up to 380 PFLOP32s (i.e. fp32 PFLOPS) of GPU-based compute from the Clouds, but that was executed as a short burst during a carefully chosen timeframe and without budgetary constraints.

We thus decided to perform a second Cloud-based run, with the additional goal of showing the available capacity during a typical workday while restricting ourselves to only the most cost-effective Cloud instance types in either spot or preemptible mode. We have also foregone the use of a dedicated setup, combining the provisioned Cloud resources with an existing resource pool. We believe than any production user would use a similar setup.

Like in the previous exercise, the chosen application was IceCube's photon propagation simulation [2], for technical (heavy use of GPU at modest IO) and scientific reasons (high impact science). IceCube follows the distributed High Throughput Computing (dHTC) paradigm and has an existing HTCondor [3] setup that regularly provisions resources from external sources, including the Open Science Grid (OSG) [4], the Extreme Science and Engineering Discovery Environment (XSEDE) and the Pacific Research Platform (PRP) [5]. We again provisioned resources from multiple public Cloud providers, and used several geographically distributed regions in each, too.

We executed the run on a Tuesday during work hours, without any coordination with or even advance notification of the Cloud providers. The observed plateau was about 15k GPUs of the most cost-effective type, which provided about 170 PFLOP32s (i.e. fp32 PFLOPS) and 64M GPU cores. The run latest about 8 hours, with the plateau being sustained for 6 hours, resulting in a total integrated compute of approximately one EFLOP32 hour.

Section 2 provides an overview of the multi-Cloud setup, including the reasoning behind choosing specific Cloud instance types and the description of the provisioned resources. Section 3 provides an analysis of the effectiveness of the setup from the application point of view. Section 4 provides a description of input data handling. And, finally, section 5 describes the science behind the simulation application as well as a summary description of the simulation code internals.

## 1.1 Related work

Running scientific workloads in the public Cloud is hardly a novel idea, nor is integration of Cloud resources in existing resource pools [1, 6-8]. This work is however novel in that it provides a measurement and cost analysis of available Cloud capacity across multiple public Cloud providers, with a focus on cost-efficient GPU resources in preemptible mode. Moreover, running an unmodified, production scientific code in such a setup is also quite unusual.

## 2 The multi-Cloud, multi-region setup

One of the main objectives of this Cloud run was to show how much spare cost-effective GPU capacity is available on a typical workday in three public Cloud providers, namely Amazon Web Services (AWS), Microsoft Azure and Google Cloud Platform (GCP), which also implied that resources had to be provisioned from Cloud regions located all over the world. We chose to use the two most cost-effective GPU types offered by each Cloud provider, based on runtimes and list cost analysis performed after the previous exercise [1]. This meant NVIDIA Tesla T4 and V100 GPUs for AWS and GCP, and NVIDIA Tesla P40 and V100 GPUs on Azure. We requested only spot instances on AWS and Azure, and preemptible instances on GCP.

The workload management system used was the existing IceCube's HTCondor system, to which we added a couple of additional scheduling and collector nodes to support the additional load. Similarly to the previous exercise, the provisioned Cloud resources were launched using spot fleets on AWS, Virtual Machine Scale Sets on Azure, and Instance Groups on GCP. For each Cloud platform we had a customized image based on CentOS Linux, containing a fully configured HTCondor worker setup, and using CVMFS [9] for software distribution. We also deployed a service instance in each Cloud region, which served both as a HTCondor collector concentrator and CVMFS cache.

The Cloud run was executed on a Tuesday in February, starting around 9:45am PST and was sustained until about 5:45pm PST. We chose to first provision T4 GPU instances only, since we expected them to be the most cost-effective ones. The other GPU types were added only after reaching an apparent plateau for the T4 GPUs, as shown in Figure 1. In the same figure you can also see how the GPU instances were distributed across the various geographic regions. The resources came from 25 Cloud regions and 20 on-prem locations distributed across the four major geographical areas.

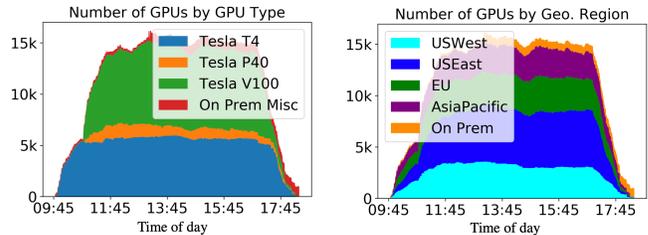

Figure 1: Number of provisioned Cloud resources, alongside on-prem resources. Left) By GPU type. Right) By geographic region.

The number of GPUs provisioned at peak during this run was much smaller than in our previous all-GPU exercise, when we managed to reach over 51k GPUs. Nevertheless, the provisioned GPUs were on average significantly more powerful, so during the plateau we reached approximately 170 PFLOP32s of compute, almost half compared to the 380 PFLOP32s of the all-GPU Cloud run, but still significantly more than the biggest XSEDE system currently deployed. And the total integrated compute exceeded one EFLOP32h, or 1000 PFLOP32 hours, with about one third coming from T4 GPUs, as seen in Figure 2. The FLOP32s represent the peak fp32 FLOPS provided by NVIDIA specs for the GPUs involved.

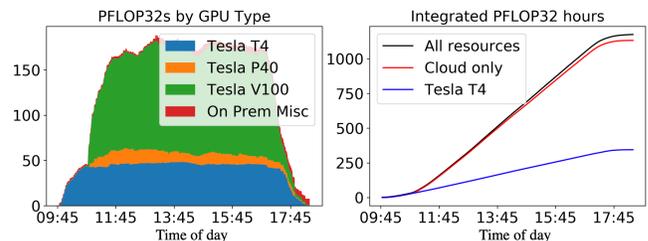

Figure 2: Total performance of provisioned Cloud resources, alongside on-prem resources, in fp32 PFLOPS. Left) Instantaneous by GPU type. Right) Integral over time.

While we are not authorized to release the precise cost of this Cloud run, we can provide an approximate value. The total cost of this Cloud run was roughly $60,000, with the T4 providing Cloud instances costing about $9,000 in total. This means that the instances providing NVIDIA T4 GPUs are about twice as cost-effective as the sum of all resources used, since they delivered approximately 30% of the integrated compute at 15% of the price. Using only T4 GPUs would have of course drastically lowered the FLOP32s being sustained during the plateau.

This run thus demonstrated that it is indeed possible to get significant cost-effective compute capability out of the public Cloud providers, if one can live with preemption and can aggregate the resources from many independent sources. Using T4 GPUs alone, one can apparently add about 40 PFLOP32s worth of compute to an existing pool for slightly more than $1,000/h, and about 170 PFLOP32s for roughly $10,000/h.

## 3 The application view of the setup

Aggregating a large amount of compute power is an interesting system administrator exercise, but did it actually allow the IceCube application to perform the desired science computation?

The provisioned Cloud resources used spot instances from AWS and Azure, and preemptible instances from GCP because they are priced at about 1/3rd the price of the on-demand equivalents. This implies that some waste had to be expected due to preemption. The used application, IceCube photon propagation simulation, does not use checkpointing, but its runtime is relatively short, from about 25 minutes for the V100 GPUs to 55 minutes for the T4 GPUs, as shown in Figure 3. On some slower GPUs provided by OSG and PRP, who also operate in preemptible mode, the application may run for up to 2 hours.

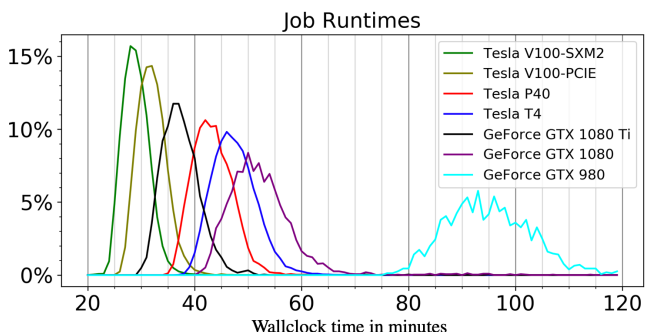

Figure 3: Observed job runtimes for the IceCube photon propagation simulation for various GPU types, in minutes.

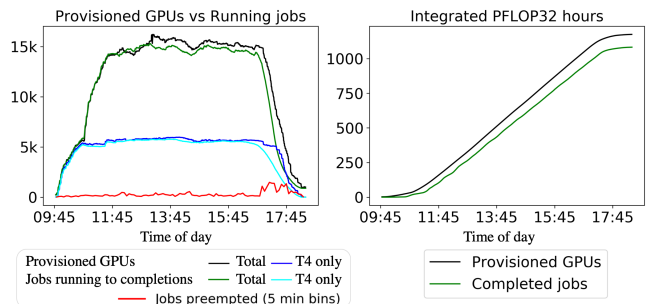

Figure 4: Difference between provisioned GPUs and jobs that ran to completion.

So, while we did observe preemptions, as shown in Figure 4, they affected only a small fraction of the jobs, and those were automatically restarted by HTCondor, resulting only in some wasted GPU cycles. Moreover, the total waste due to preemption was less than 10%, as shown in the same figure. Note that there was some waste incurred during the rampdown sequence, too, as we did not always de-provision exactly at job termination boundary. Given that spot and preemptible Cloud instances are billed at only a fraction of the cost of more reliable on-demand instances, the observed waste is a very cost-effective tradeoff.

Observing the number of jobs completed on the different resources shows that using FLOP32s as a metric does provide a valid comparison. About a third of all the jobs ran on the NVIDIA Tesla T4 instances, as can be seen from Figure 5, which is comparable to the fraction of integrated FLOP32s that those instances provided, as shown in Figure 2. The observation that using just T4 GPUs would be twice as cost-effective than using all three types of Cloud GPUs, at the expense of speed of progress, is thus still valid.

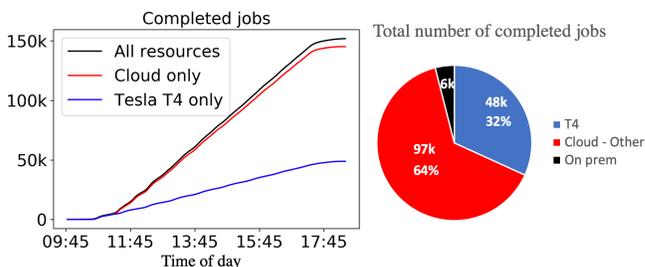

Figure 5: Number of completed IceCube jobs.

It is also worth noting that during this Cloud run we produced the output of 151k jobs, which is about 50% more than the 101k that were produced during the previous exercise. While the previous exercise did reach a higher peak, the total value for the IceCube science was greater this time.

## 4 Input file handling

The IceCube photon propagation application expects an input file containing a set of photons to propagate, which is generated asynchronously by another process in the workflow. We had many such files ready to be processed on IceCube servers located at the University of Wisconsin – Madison (UW), each about 45 MB in size. All the files were accessible through a Web portal, using the HTTP protocol, which was tested as being able to deliver files at up to 100 Gbps.

Each job during this Cloud run would fetch an input file from UW using a command line tool, typically aria2, store it on local disk and then launch the photon propagation application process. We logged the amount of time it took for each job to fetch the input data and are glad to report that for most jobs it took less than 10 seconds, as can be seen from Figure 6. Given that typical job runtime was in the 1.5k to 3.5k seconds range, the overhead of fetching files was negligible. The total needed throughput was about 4 Gbps, with a slightly higher peak during the initial rampup. Cloud networking is thus fast enough, even from remote regions, to allow for both quick bursting and sustained compute at large scales for applications with modest input data needs. It should also be noted that incoming networking is not a billable item in the three public Cloud providers used.

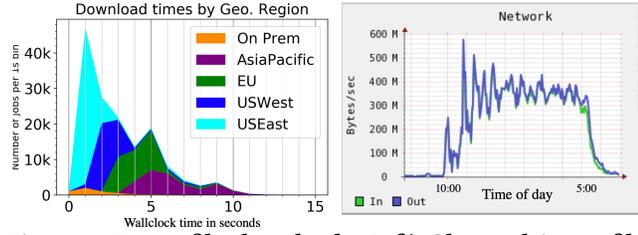

Figure 6: Input file downloads. Left) Observed input file download times, in seconds. Right) Observed throughput, as measured on the server.

For this exercise, we still first uploaded the output files into Cloud native storage and then fetched them asynchronously back to UW. We did this since we did not have an object store available at UW in time for the Cloud run. We will likely explore direct upload to UW in a follow up exercise.

## 5   IceCube science motivation

The IceCube Neutrino Observatory [10] is the world's premier facility to detect neutrinos with energies above 1 TeV and an essential part of multi-messenger astrophysics. IceCube is composed of 5160 digital optical modules (DOMs) buried deep in glacial ice at the geographical south pole. Neutrinos that interact close to or inside of IceCube produce secondary particles, often a muon. Such secondary particles produce Cherenkov (blue as seen by humans) light as they travel through the highly transparent ice. Cherenkov photons detected by DOMs can be used to reconstruct the direction and energy of the parent neutrino.

Since the detector is built into a naturally existing medium, i.e. glacial ice, there was a priori only limited information regarding the optical properties of the detector, so a lot of simulation data is needed to properly calibrate the employed instruments. The optical properties of the glacial ice greatly affect the pointing resolution of IceCube. Improving the pointing resolution has two effects in this case: greater chance to detect astrophysical neutrinos and better information sent to the community. While IceCube can detect all flavors and interaction channels of neutrinos, about two-thirds of the flux reaching IceCube will generate a detection pattern with a large angular error, see Figure 7. In the same figure you can also see that this angular error is mostly driven by systematic effects. Similarly, different optical models have a great effect on the reconstructed location of an event on the sky. The comparatively minute field of view of partner observatories and telescopes requires IceCube to provide as accurate as information as possible.

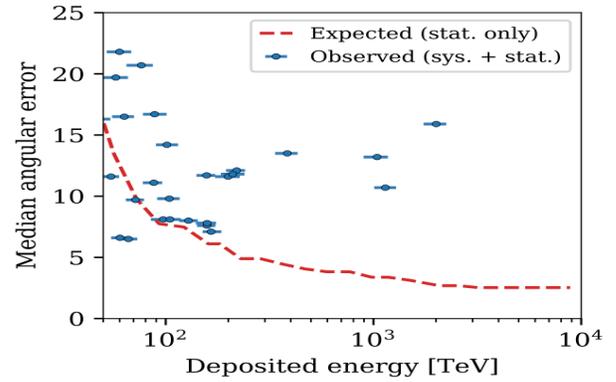

Figure 7: Impact of the IceCube detector calibration on science results.

The most computationally intensive part of the IceCube simulation workflow is a photon propagation code, and that code can greatly benefit from running on GPUs. The algorithm follows these steps. Initially a set of photons is created along the path of charged particles produced in the neutrino interaction or from in-situ light sources used for calibration. Once the location and properties of the photons have been determined, they are added to a queue. A thread pool is created depending on the possible number of threads, typically one to several threads per GPU core, with the exact mapping depending on the specific vendor and architecture. Each thread takes a photon out of the queue and propagates it. During the propagation, the algorithm will first determine the absorption length of the photon, i.e. how long the photon can travel before being absorbed. Then the algorithm will determine the distance to the next scatter. The photon is now propagated the distance of the next scatter. After the propagation, a check is performed to test whether the photon has reached its absorption length or intersected with an optical detector along its path. If the photon does not pass these checks, the photon is scattered, i.e. a scattering angle and a new scattering distance are determined, and the cycle repeats. Once the photon has either been absorbed or intersected with an optical detector, its propagation is halted and the thread will take a new photon from the queue.

The IceCube photon propagation code is distinct from others, e.g. Nvidia OptiX in that it is purpose-built. It handles the medium, i.e. glacial ice and the physical aspects of photon propagation in great detail. The photons will traverse through a medium with varying optical properties. The ice has been deposited over several hundreds of thousands of years. Earth's climate changed significantly during that time and imprinted a pattern on the ice as a function of depth. In addition to the depth-dependent optical properties the glacier has moved across the Antarctic continent and has undergone other unknown stresses. This has caused layers of constant ice properties, optically speaking, to be tilted and to have anisotropic optical properties.

## 6 Conclusions

This paper presents our experience in expanding an existing production HTCondor pool by several orders of magnitude for the duration of a workday, using cost-effective GPU instances in preemptible mode gathered from the three major Cloud providers, namely Amazon Web Services, Microsoft Azure and the Google Cloud Platform. The chosen community was IceCube, and the application their photon propagation simulation, both for technical (heavy use of GPU at modest IO) and scientific reasons (high impact science). The needed input data was fetched straight from IceCube's Web servers at UW Madison, with minimal impact on overall job efficiency.

We managed to provision about 15k instances, a mix of NVIDIA Tesla T4, P40 and V100 GPUs, corresponding to around 170 PFLOP32s. We executed this run on a Tuesday in February and ran for a whole workday, sustaining a plateau which was close to the peak for about 6 hours. The observed waste due to preemption was less than 10%. The total cost of this exercise was about $60,000, or slightly less than $10,000 per hour.

It is also worth noting that we provisioned and sustained about 5.5k instances having T4 GPUs, for about 45 PFLOP32s, at the total cost of about $9,000, or slightly more than $1,000 per hour. The T4-providing instances completed almost one third of all IceCube jobs, making them about twice more cost-effective than the sum of all the resources we provisioned, if slower progress is acceptable.

We made no special arrangements or had long-term commitments with the Cloud providers to achieve this. We are thus confident we could repeat such runs on a regular basis, possibly several times a week, if we had the funding to do so.

## ACKNOWLEDGMENTS

This work was partially funded by US National Science Foundation (NSF) under grants OAC-1941481, MPS-1148698, OAC-1841530, OPP-1600823, OAC-190444 (as a subaward through Internet2's E-CAS project), and OAC-1826967.